\documentclass{ptapap}

\usepackage{amssymb}

\author{James Sikora}[Queens,RMC]
\author{Gregg Wade}[RMC]
\author{Jason Rowe}[Bishops]
\affil[Queens]{Dept. of Physics, Queen's University\\
  Kingston, ON, Canada}
\affil[RMC]{Dept. of Physics, Royal Military College of Canada\\
  Kingston, ON, Canada}
\affil[Bishops]{Dept. of Physics and Astronomy, Bishop's University\\
  Sherbrooke, QC, Canada}

\title{Deciphering the Surprising Variability of A-type Stars Revealed by Kepler}

\begin{document}

\maketitle

\begin{abstract}

A recent analysis of high precision photometry obtained using the \emph{Kepler} spacecraft has revealed 
two surprising discoveries: (1) over 860 main sequence A-type stars -- approximately $40\,\%$ of those 
identified in the \emph{Kepler} field -- exhibit periodic variability that may be attributable to 
rotational modulation by spots and (2) many of their light curves indicate the presence of a mysterious 
and characteristic power spectral feature. We have been carrying out an ongoing analysis designed to 
expand upon these discoveries and to provide a possible explanation for the unusual power spectral 
features. In the following, we will put these recent discoveries into context as well as discuss the 
preliminary findings yielded by our analysis of the \emph{Kepler} light curves.

\end{abstract}

\sloppy

\section{Introduction}

In general, main sequence (MS) A-type stars are not known to be significantly active. They do not exhibit 
the high-energy, X-ray emitting flares and coronal mass ejections that are produced by low-mass MS stars. 
Nor are they hot enough to emit dense, high-velocity line-driven winds such as those associated with 
early B- and O-type MS stars. As such, they are often defined by their intermediate mass or ``tepid" 
surface temperatures \citep[e.g.][]{Landstreet2003_symp} in part as a way of distinguishing them from 
their more active low- and high- mass counterparts. However, with the availability of space-based 
photometric observations such as those obtained by \emph{Kepler}, it has become evident that a 
significant fraction of MS A-type stars do exhibit low-amplitude variability; in many cases, this 
phenomena currently defies an explanation.

There are various mechanisms which can induce periodic brightness variations in MS A-type stars. 
Pulsations are one example, which can occur for stars located within well-studied instability strips on 
the Hertzsprung-Russell diagram. This kind of photometric variability occurs at both low frequencies 
($\lesssim5\,{\rm d^{-1}}$, e.g. $\gamma$ Dor stars) and high frequencies ($>5\,{\rm d^{-1}}$, e.g. 
$\delta$ Scuti stars) with amplitudes $\lesssim0.5\,{\rm mag}$. The presence of surface structures (e.g. 
chemical spots) may also result in brightness variations as a consequence of flux redistribution in 
combination with a star's rotation. These so-called $\alpha^2$ CVn variable stars exhibit photometric 
amplitudes $\sim1-10\,{\rm mmag}$ with frequencies $\lesssim3.5\,{\rm d^{-1}}$ where the upper limit is 
based on the critical rotational frequency of a typical MS A-type star.

It has long since been established that the origin of the spots associated with $\alpha^2$ CVn stars is 
related to the presence of strong ($B\gtrsim300\,{\rm G}$), stable, and organized magnetic fields and 
that these objects are in fact equivalent to the class of chemically peculiar magnetic Ap/Bp stars. More 
recently, a new class of magnetic MS A-type star has emerged following the detection of ultra-weak 
fields ($B\lesssim1\,{\rm G}$) on the surfaces of Vega and Sirius A \citep{Lignieres2009,Petit2011}. 
Given the essentially one-to-one correlation between $\alpha^2$ CVn variables and magnetic Ap/Bp stars, 
it is perhaps unsurprising that the detection of a magnetic field on Vega was soon followed by the 
detection of spots on its surface \citep{Bohm2015}. These findings may well suggest that a large fraction 
of MS A-type stars -- much larger than the $10\,\%$ incidence rate of Ap/Bp stars 
\citep[e.g.][]{Wolff1968,Auriere2007} -- host ultra-weak fields on their surfaces and exhibit 
low-amplitude photometric variability \citep{Petit2011a}.

In 2011, Balona carried out an analysis of \emph{Kepler} light curves associated with $\sim1,600$ 
A-type stars and reported that a large fraction of them exhibit sharp, low-frequency peaks in their 
periodograms that are consistent with rotational modulation with periods of a few days 
\citep{Balona2011}. This study was expanded upon with the release of additional \emph{Kepler} data 
\citep{Balona2013} culminating in the identification of 887 rotationally modulated A star light curves 
-- $44\,\%$ of those found within the original \emph{Kepler} field. It has been suggested that this 
variability may be produced by spots similar to those detected on Vega \citep{Bohm2015,Blazere2016a}.

\begin{figure}
\includegraphics[width=\textwidth]{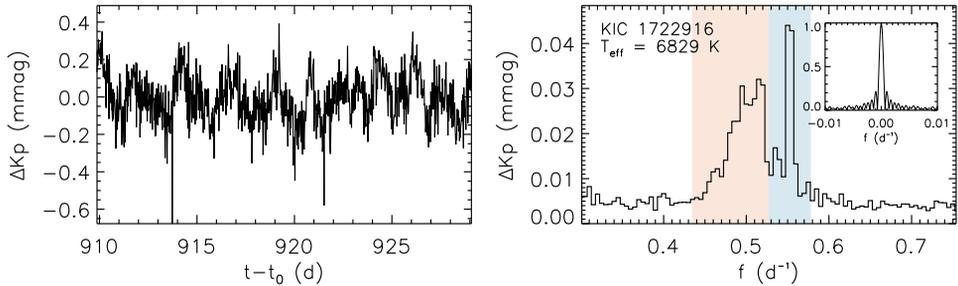}
\caption{An example of a comparatively cool star ($T_{\rm eff}=6\,829\,{\rm K}$) observed by Kepler (KIC 
1722916) that was found to exhibit the peculiar periodogram feature. \emph{Left:} A subsample of the full 
light curve. \emph{Right:} A region of the binned Lomb-Scargle periodogram associated with the star's 
light curve. The peculiar periodogram feature is defined by the presence of a diffuse peak (highlighted in 
pink) and a narrow peak (highlighted in blue) with the former occuring at slightly higher frequencies 
($\gtrsim0.05\,{\rm d^{-1}}$) relative to the latter. The inset plot shows the light curve's spectral 
window.}
\label{fig:PPF_ex}
\end{figure}

In addition to this discovery, \citet{Balona2013} reported the widespread detection of characteristic 
peculiar features in the periodograms. These ``peculiar periodogram features" consist of a diffuse peak 
accompanied by a narrow peak at slightly higher frequency (Fig. \ref{fig:PPF_ex}). 
\citet{Balona2013,Balona2014} speculated that the peculiar periodogram features are caused by the 
combination of two phenomena: (1) the diffuse peaks are produced by the presence of spots in the 
photospheres of each A star, where the broadening is a consequence of differential rotation; (2) the 
narrow peaks are produced by a hot Jupiter orbiting the A star with a period only slightly shorter than 
the star's rotational period ($\sim1-3\,{\rm d}$). In the following, we propose an alternative 
explanation for the origin of these peculiar periodogram features and summarize some of the preliminary 
findings of our ongoing study.

\section{Sub-surface Structures}\label{sect:hypothesis}

Consider a light curve having variability that can be attributed to the presence of inhomogeneous 
structures on the rotating star's surface. In the case of light curves obtained by \emph{Kepler}, the 
measurements span long timescales such that any broadening of the periodogram peaks occuring at the 
star's rotational period can be interpreted in two ways: (1) it is caused by structures having 
relatively short lifetimes (in comparison to the observing timescale) or (2) it is caused by 
differential rotation in which the rotational frequencies of the structures depend on their latitudes. 
Narrow peaks having widths comparable to the periodogram's frequency resolution are indicative of 
surface structures that are stable (both in brightness and in frequency) over the observing timescale. 
Given that the peculiar periodogram features consist of both broad and narrow peaks that are correlated 
in frequency, it is difficult to attribute their origin to any currently understood surface structures.

We propose that the broad and narrow peaks defining the peculiar periodogram features are produced by 
inhomogeneous structures localized within two distinct regions. In this scenario, the narrow 
periodogram peaks are produced on or near the (presumably) rigidly rotating stellar surface. On the 
other hand, the broad peaks are produced by similar structures forming at lower depths near a 
sub-surface convective-radiative boundary. This hypothesis immediately yields two predictions: we 
expect that, as the depth of this convective-radiative boundary decreases with decreasing effective 
temperature, both (1) the incidence rate of the peculiar periodogram feature and (2) the amplitude of 
the broad peaks will decrease in stars with lower $T_{\rm eff}$.

At this point, our hypothesis has not been fully developed and has several aspects that are not 
constrained. For instance, it may be the case that it is the layers closer to the surface that are 
undergoing differential rotation while the sub-surface structures are rotating rigidly, rather than the 
other way around. It is also possible that, assuming that the broadening is produced by differential 
rotation and not by short structure lifetimes, the differential rotation is radial rather than 
latitudinal. One important question that has yet to be answered is to what depth we should expect the 
hypothesized sub-surface structures to be visible. This requires a number of assumptions regarding the 
nature of the structures themselves. Specifically, it is unknown how the brightness of each structure 
might vary relative to the surrounding material or how far they might extend -- both horizontally and 
radially.

\section{Incidence of the Peculiar Periodogram Features}

\citet{Balona2013} identified 129 MS A-type stars which exhibit the peculiar feature in the periodograms 
calculated from their \emph{Kepler} light curves. Using the Kepler Input Catalogue \citep{Brown2011}, 
it was concluded that the occurence of the feature did not correlate with any of the stars' derived 
fundamental parameters. In particular, no significant trend was found in relation to their effective 
temperatures. Although these findings are inconsistent with the first prediction discussed in 
Section \ref{sect:hypothesis}, it is possible that the predicted changes in sub-surface convection zones 
in stars with $7.5\leq T_{\rm eff}\leq10\,{\rm kK}$ are subtle and require larger sample sizes to be 
discerned. Moreover, since \citet{Balona2013} published their results, additional \emph{Kepler} data has 
been made available, which may allow for previously unidentified peculiar periodogram features to be 
found within the sample of MS A-type stars.

\begin{figure}
\begin{center}
\includegraphics[width=0.8\textwidth]{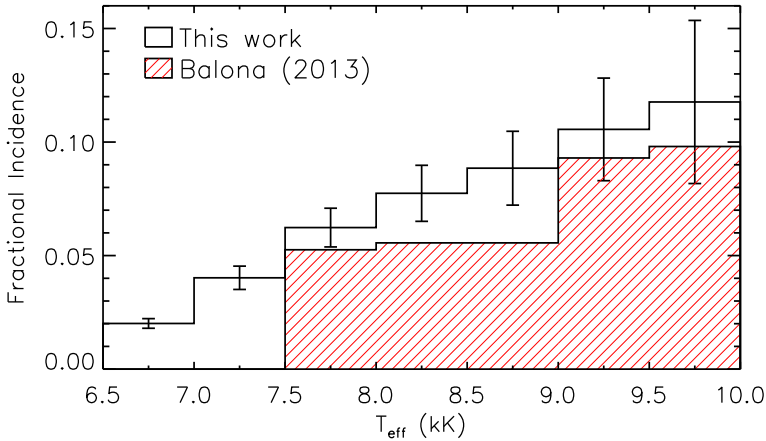}
\caption{The fractional incidence rates of Kepler stars exhibiting the peculiar periodogram feature as 
a function of $T_{\rm eff}$. The red hatched distribution is that reported by \citet{Balona2013} while 
the white distribution was found in this study. It is evident that the fraction of stars exhibiting the 
peculiar feature decreases with decreasing $T_{\rm eff}$ from $\sim12\,\%$ in early A-type to 
$\sim2\,\%$ in mid F-type stars.}
\label{fig:Teff_incid}
\end{center}
\end{figure}

We endeavoured to carry out an expanded analysis of the \emph{Kepler} light curves associated with the 
observed MS A-type stars in order to confirm whether any correlations in the incidence rate of the 
peculiar periodogram feature with $T_{\rm eff}$ (and other parameters) could be found. We extended 
the effective temperature range used by \citet{Balona2013} to $6.5\leq T_{\rm eff}\leq10\,{\rm kK}$, 
thereby including cooler stars, which resulted in a total sample size of 8,377 stars -- an increase of 
more than 6,000 stars from the original sample. In order to analyze this much larger sample, we opted to 
employ a convolution neural network (CNN). These machine learning algorithms have been applied to a wide 
range of signal-processing problems in which specific features in time-series data need to be efficiently 
and accurately classified \citep[e.g.][]{George2017}. We generated $\sim100,000$ simulated periodograms 
containing isolated sharp peaks, isolated broad peaks, and combined sharp and broad peaks characteristic 
of the peculiar periodogram features (Fig. \ref{fig:PPF_ex}). This data set was then used to train our 
CNN to classify each of these features with an accuracy $\gtrsim93\,\%$; the \emph{Kepler} 
periodograms associated with the F- and A-type stars were then applied to the trained CNN. The resulting 
incidence rate of these features as a function of $T_{\rm eff}$ is shown in Fig. \ref{fig:Teff_incid}.

\section{Conclusions}

Based on our preliminary findings, it is evident that the peculiar periodogram features first identified 
by \citet{Balona2013} do appear more frequently in early A-type MS stars ($\sim12\,\%$) compared to 
cooler mid F-type stars ($\sim2\,\%$). This is broadly consistent with the hypothesis that the features 
are associated with sub-surface convection zones, which tend to increase in depth with decreasing 
effective temperature; however, a more comprehensive comparison with the predictions of stellar 
structure models needs to be carried out.

\bibliographystyle{ptapap}
\bibliography{ptapapdoc_mod}

\end{document}